# Spontaneous Emission Spectrum in Double Quantum Dot Devices

Toshimasa Fujisawa[1,2], Tjerk H. Oosterkamp[1], Wilfred G. van der Wiel[1], Benno W. Broer[1], Ramón Aguado[1], Seigo Tarucha[2,3] and Leo P. Kouwenhoven[1].

[1] *Department of Applied Physics and DIMES, Delft University of Technology, P.O. Box 5046, 2600 GA Delft, The Netherlands.*
[2] *NTT Basic Research Laboratories, 3-1, Morinosato-Wakamiya, Atsugi, Kanagawa, 243-0198, Japan.*
[3] *Department of Physics, University of Tokyo, 7-3-1 Hongo, Bunkyo-ku, Tokyo 113-0033, Japan.*

A double quantum dot device is a tunable two-level system for electronic energy states. A dc electron current directly measures the rates for elastic and inelastic transitions between the two levels. For inelastic transitions energy is exchanged with bosonic degrees of freedom in the environment. The inelastic transition rates are well described by the Einstein coefficients, relating absorption with stimulated and spontaneous emission. The most effectively coupled bosons in the specific environment of our semiconductor device are acoustic phonons. The experiments demonstrate the importance of vacuum fluctuations in the environment for little circuits of coherent quantum devices.

Electronic quantum devices explore quantum mechanical properties of electrons confined to small regions in a solid by means of modern fabrication techniques. Existing devices include semiconductor resonant tunneling diodes (1) (based on quantum mechanical confinement), superconducting Josephson junction circuits (2) (based on macroscopic phase coherence), metallic single electron transistors (3) (based on quantization of charge), and molecular electronic devices (4). The principle of operation in circuits of these devices is based on controlling energy states, for instance, by means of an external (gate) voltage. One source for unwanted transitions and errors is always the thermal energy from a non-zero temperature. However, even at zero temperature vacuum fluctuations in the environment can give rise to transitions between states of non-equal energy by spontaneous emission of an energy quantum. Such inelastic transitions cause errors in many proposed schemes for quantum circuits. We have studied inelastic transitions in a fully-controllable, two-level quantum system realized in a double quantum dot device. We can relate the transition rates involving emission to absorption rates by the Einstein coefficients over our full energy and temperature range. At our lowest temperature we directly measure the energy-dependent rate for spontaneous emission. In our specific semiconductor device this energy is emitted into the environment formed by acoustic phonons.

Our double quantum dot (Fig. 1A) is fabricated in the two-dimensional electron gas (2DEG) of an AlGaAs/GaAs semiconductor heterostructure (5). The source and drain are large 2DEG regions which serve as leads for contacting current and voltage wires. The two dots, $L$ and $R$, are separated from each other and from the leads, by potential barriers induced by negative voltages applied to the three metallic gates. Tunneling between the different regions is sufficiently strong to detect current, but weak enough such that the number of electrons in each dot is a well-defined integer. The energy states in such fully confined regions are discrete, 0D-states; resembling discrete atomic states (6, 7). The discrete energies include contributions from single-electron charging energies, arising from Coulomb interactions, and from quantum-mechanical confinement. The lowest energy state for one additional electron in the left dot is labeled in Fig. 1B-D as $E_L$ and similarly $E_R$ for the right dot. Fig. 1C illustrates the resonance condition, $E_L = E_R$, in which case an electron can tunnel elastically from an occupied state in the source via $E_L$ and $E_R$ to an empty state in the drain. Such tunneling sequences of single electrons are regulated by the Coulomb charging energies (3, 7). When the two states are not aligned, $E_L \neq E_R$, only inelastic transitions are allowed for which some energy needs to be exchanged with the environment. A measured off-resonance current, therefore, directly provides information about the coupling between electrons on the dots to degrees of freedom in the environment. The inelastic rates can be analyzed with well-developed methods in quantum optics (8, 9).

Figure 1E shows a typical current spectrum versus $\varepsilon \equiv E_L - E_R$ at our lowest lattice temperature $T = 23$ mK (10). The gate voltages $V_{GR}$ and $V_{GL}$ are swept simultaneously such that the respective energies are like in Fig. 1B-D; that is, $\varepsilon = 0$ occurs in the middle between the Fermi energies of source and drain, $\mu_S$ and $\mu_D$, and $|\varepsilon| = eV_{SD}$ is maximal corresponds to having the states $E_L$ and $E_R$ aligned to one of the Fermi energies. To analyze the large asymmetry, we decompose the total current $I_{tot}(\varepsilon) = I_{el}(\varepsilon) + I_{inel}(\varepsilon > 0)$ into a symmetric part $I_{el}(\varepsilon) = I_{el}(-\varepsilon)$ (dashed curve) and the remaining asymmetric part $I_{inel}(\varepsilon > 0)$ (dotted-dashed curve). At $T = 0$, $I_{el}(\varepsilon)$ is due to elastic tunneling and has a Lorentzian lineshape $I_{el}(\varepsilon) = I_{el,\max} w^2/(w^2 + \varepsilon^2)$ (11). The full width at half maximum (FWHM), $2w$, can be tuned by the central gate voltage $V_{GC}$ roughly from 4 to 20 $\mu$eV. From measurements of $I_{el}(\varepsilon)$ at positive and negative $V_{SD}$ it is possible to extract values for the tunnel couplings $\Gamma_L$, $\Gamma_R$ and $T_c$ (11, 12).

The remaining current, $I_{inel}(\varepsilon > 0)$, which is non-zero only for $\varepsilon > 0$, is due to inelastic tunneling. In Fig. 1E, $I_{inel}$ is non-zero over an energy range of $\sim$100 $\mu$eV; this despite that the thermal energy $kT$ (23 mK) = 2 $\mu$eV is much smaller. (The irregular fine structure is discussed below.) In general we find that $I_{inel}$ vanishes when one of the levels, $E_L$ or $E_R$, crosses one of the two Fermi energies. In the specific case of Fig. 1E, $E_L$ and $E_R$ cross the Fermi energies simultaneously, implying that $I_{inel}$ is cut off at $\varepsilon = eV_{SD}$. Below this cut off, the value of $I_{inel}$ was not influenced by the value of $V_{SD}$ (13). For $T = 0$, we can write the condition for a non-zero inelastic current as $\mu_S > E_L > E_R > \mu_D = \mu_S - eV_{SD}$. The amount of inelastic current depends on the transition rates as: $I_{inel}(\varepsilon) = e(\Gamma_L^{-1} + \Gamma_i^{-1}(\varepsilon) + \Gamma_R^{-1})^{-1}$. When the inelastic rate $\Gamma_i(\varepsilon)$ from $E_L$ to $E_R$ is much smaller than the rates through the outer barriers, this reduces to $I_{inel}(\varepsilon) = e\Gamma_i(\varepsilon)$.

The effect of a non-zero temperature on the current is shown in Fig. 2A. A higher temperature $T$, enhances $I_{tot}$ on both the emission ($\varepsilon > 0$) and the absorption ($\varepsilon < 0$) side. The absorption spectrum shows an exponential temperature dependence, $e^{\varepsilon/kT}$ (dashed lines) for absolute energies larger than the elastic current measured at 23 mK, that is $|\varepsilon| > w$.

To analyze the temperature dependence, we assume boson statistics for the degrees of freedom in the environment. The average occupation number $\langle n \rangle$ of environmental modes at energy $\varepsilon$ is given by the Bose-Einstein distribution function: $\langle n \rangle = 1/(e^{\varepsilon/kT} - 1)$. The rates for absorption, $W_a$, and emission, $W_e$, can be expressed very generally by $W_a = B_a\rho$ and $W_e = A + B_e\rho$, where the Einstein coefficients stand for spontaneous emission ($A$), stimulated emission ($B_e$) and absorption ($B_a$), and $\rho$ is the energy density (8). From the Einstein relations, $B_a = B_e = A\langle n \rangle/\rho$ (8), we obtain:

$$\Gamma_i(\varepsilon < 0) = W_a(\varepsilon) = \langle n \rangle A(-\varepsilon) \quad (1)$$
$$\Gamma_i(\varepsilon > 0) = W_e(\varepsilon) = (\langle n \rangle + 1)A(\varepsilon)$$

To test whether the inelastic current follows emission and absorption statistics, we calculate the full current spectrum from Eqs. 1. First, we obtain the spontaneous emission rate from $A(\varepsilon) = I_{inel}(\varepsilon > 0, T = 23 \text{ mK})/e$. The trace at 23 mK is effectively at zero temperature for $\varepsilon \gg 2$ $\mu$eV since then $\langle n \rangle \ll 1$. The emission current at higher temperatures follows from $I_{inel}(\varepsilon > 0, T) = e(\langle n \rangle + 1)A(\varepsilon)$, whereas the absorption current follows from $I_{inel}(\varepsilon < 0, T) = e\langle n \rangle A(-\varepsilon)$. The reconstructed current spectrum is shown in Fig. 2B. The central part of the curves ($|\varepsilon| < 10$ $\mu$eV) is kept blank since Eq. 1 does not include the $T$-dependence of $I_{el}$. The calculated current reproduces the measured current well up to 200 mK. Even the small step-like feature seen at $\varepsilon \sim 30$ $\mu$eV is reflected by a shoulder-like feature at $\varepsilon \sim$ -30 $\mu$eV in the measured and in the calculated absorption spectra (indicated by arrows). For $T > 200$ mK the measured current significantly exceeds the calculated current, which is probably due to thermally excited electrons (Eq. 1 only describes the $T$-dependence of the environment. The thermal excitations in the electron leads are not included.) Further confirmation of the applicability of the Einstein relations to our quantum dot system follows from the prediction: $[I_{inel}(\varepsilon > 0) - I_{inel}(\varepsilon < 0)]/eA(|\varepsilon|) = [W_e - W_a]/A = 1$, which is valid independent of temperature. Fig. 2C shows a plot of the normalized rates, $W_a/A$ and $W_e/A$, versus $kT/|\varepsilon|$ for various $\varepsilon$ and $T$ up to 200 mK. The measured data closely follow the prediction $[W_e - W_e]/A = 1$; that is, the normalized rates, $W_a/A$ and $W_e/A$, differ by one over the temperature range $T < 200$ mK without fitting any parameter.

The inelastic rate for a two-level system coupled to a bosonic environment at $T = 0$ is expected to have a $T_c^2$ dependence (14, 15). Still without identifying the bosonic environment, we can test this dependence on the elastic tunnel coupling $T_c$ between the two dots. Figure 3A shows that the inelastic current clearly increases with $T_c$. For the largest coupling we obtain a saturation where the elastic current peak can no longer be distinguished. By fitting the elastic current part to a Lorentzian lineshape (11) we can obtain rough estimates for $T_c$ as long as the current is less than the saturation value. We find that with these fitted values, the inelastic current scales as $T_c^\alpha$ with an exponent $\alpha = 2.5 \sim 3$, maybe somewhat larger than expected. Figure 3B shows the effect of the increased coupling on the symmetric part of the current at low temperature. For small tunnel coupling, we always obtain Lorentzian lineshapes. For increasing couplings, the data still fits to a Lorentzian tail on the absorption side. However, we generally find significant deviations for small $\varepsilon$, implying that for large coupling the elastic and inelastic rates can become of the same order. This may form a significant limitation for the coherence time in coupled quantum devices (16).

The importance of fluctuations in the environment on electron tunneling through quantum devices has been recognized for a long time. Environmental studies on Coulomb blockade devices have only discussed effects due to absorption (3). For emission it is required that electrons are first pumped to a higher energy state. This has recently been done in a superconducting Cooper pair transistor under microwave irradiation (17). In the case of a double dot, pumping occurs when $E_L > E_R$ and an electron tunnels in from the left reservoir to $E_L$. A double dot thus offers a unique two-level system that is pumped by a dc voltage without inducing heating currents. It is therefore possible to reach an out-of-equilibrium situation so close to $T = 0$ that vacuum fluctuations become the main source for generating electron transport.

To identify whether photons, plasmons, or phonons form the bosonic environment, we measured spontaneous emission spectra while placing the double dot in different electromagnetic environments. In the regime 10-100 $\mu$eV, the typical wavelengths are 1-10 cm for photons and 0.3 to 30 cm for 2DEG plasmons. We have tested the coupling to the photonic environment by placing the sample in microwave cavities of different size (18). To check the coupling to plasmons, we have measured different types of devices with largely different dimensions of the 2DEG leads, gate pads, and bonding wires. Both types of variation had no effect at all on the emission spectra; even the finestructure was reproduced.

The third option of acoustic phonons is the most likely possibility (19). Phonon emission rates have been calculated for single dots (20). For a double dot system, we can obtain the general energy dependence (15). For a deformation potential we expect a rate dependence of $\varepsilon^{D-2}$ ($\varepsilon$ for 3D phonons and constant for 2D phonons) and for piezo-electric interaction of $\varepsilon^{D-4}$ ($1/\varepsilon$ for 3D phonons and $1/\varepsilon^2$ for 2D surface acoustic waves) (21). In Fig. 3C we compare traces measured on two different types of devices. Here, the emission current is plotted versus $\varepsilon$ on a log-log scale. Ignoring the bumps, we find an energy dependence between $1/\varepsilon$ and $1/\varepsilon^2$. This implies that the dominant emission mechanism is the piezo-electric interaction with 2D or 3D acoustic phonons. Note that a $1/\varepsilon$ or $1/\varepsilon^2$ dependence should be avoided in coherent devices, since the inelastic rate becomes large near resonance ($\varepsilon \sim 0$) (16).

The bumps observed in both type of devices suggest the existence of resonances, for instance, due to a finite size in the phonon environment. The bumps are particularly clear in the derivative of the current to energy (dotted curves in Fig. 3A). The large bump in Fig. 3A at $\varepsilon = 30$ $\mu$eV corresponds to a frequency of $f = \varepsilon/h = 7.3$ GHz. For 3D phonons this yields a wavelength $\lambda^{3D} = s^{3D}/f = 640$ nm ($s^{3D} = 4800$ m/s is the 3D sound velocity), whereas for 2D surface acoustic waves $\lambda^{2D} = s^{2D}/f = 380$ nm ($s^{2D} = 2800$ m/s). These wavelengths both, more or less, fit with the dimensions of the two quantum dot devices. We have not yet been able to control these resonance by studying devices with a variety of gate dimensions. However, we believe that it is possible to gain control over the phonon environment by making 3D phonon cavities in hanging bridges (22) or by creating a 2D phonon bandgap using a periodic gate geometry (23).

a rectangular cavity ($22 \times 19 \times 8$ mm$^3$) with resonance frequency of about 40 $\mu$eV.

[19] Coupling to optical phonons is efficient only at much larger energies.

[20] U. Bockelmann, *Phys. Rev. B* **50**, 17271 (1994).

[21] For phonons the inelastic rate at $T = 0$ can be written as $\Gamma_i \sim (T_c/\varepsilon)^2 J(\varepsilon/\hbar)$ where $J(\varepsilon) \sim \frac{c^2(\varepsilon)}{\varepsilon} g(\varepsilon)$ is the spectral function (14). The phonon density of states $g(\varepsilon) \sim \varepsilon^{D-1}$, such that $\Gamma_i \sim T_c^2 c^2(\varepsilon) \varepsilon^{D-4}$. When we neglect possible fine structure in $c(\varepsilon)$, we have for deformation potential $c \sim \varepsilon$ and thus $\Gamma_i \sim \varepsilon^{D-2}$, while for piezo-electric interaction $c \sim$ constant and thus $\Gamma_i \sim \varepsilon^{D-4}$.

[22] A. N. Cleland and M. L. Roukes, *Nature*, **392**, 160 (1998).

[23] J. M. Shilton, et al., *J. Phys.: Cond. matter* **8**, L531 (1997).

[24] We thank M. Devoret, L. Glazman, S. Godijn, Y. Hirayama, J. Mooij, Yu. Nazarov, Y. Tohkura, N. Uesugi, M. Uilenreef, and N. van der Vaart for help and discussions. Supported by the Dutch Foundation for Fundamental Research on Matter (FOM), L.P.K. by the Royal Netherlands Academy of Arts and Sciences.

FIG. 1. (**A**) Double quantum dot device defined in the 2DEG of a GaAs/AlGaAs hetero structure by focused ion beam implantation. The narrow channel connects the large 2D source and drain leads. Negative voltages ($V_{GL}$, $V_{GC}$, and $V_{GR}$) applied to the metal gates ($G_L$, $G_C$, and $G_R$; widths are 40 nm) induce three tunable tunnel barriers in the wire. The two quantum dots, $L$ and $R$, respectively, contain $\sim$15 and $\sim$25 electrons; charging energies are $\sim$4 and $\sim$1 meV; and the measured average spacing between single-particle states are $\sim$0.5 and $\sim$0.25 meV. (**B**, **C**, and **D**) Energy diagrams (vertical axis) along the spatial axis through the dots (horizontal axis) for the tunnel situations: absorption, elastic and emission. Thick vertical lines denote tunnel barriers. The continuous electron states in the leads are filled up to the Fermi energies $\mu_S$ and $\mu_D$. The external voltage $V_{sd}$ between leads opens a transport window of size: $eV_{sd} = \mu_S - \mu_D$. The energy, $\varepsilon \equiv E_L - E_R$, is defined as the difference between the topmost filled discrete-state of the left dot, $E_L$, and the lowest discrete-state for adding an extra electron to the right dot, $E_R$. (The inter-dot capacitance prevents that $E_L$ and $E_R$ are simultaneously occupied.) An elastic current can flow when $\varepsilon = 0$, otherwise a non-zero current requires absorption ($\varepsilon < 0$) or emission of energy ($\varepsilon > 0$). $T_c$ is the tunnel coupling and $\Gamma_i$ is the inelastic rate between the two dots. $\Gamma_L$ and $\Gamma_R$ are the tunnel rates across the left and the right barriers. (**E**) Typical measurement of the current (solid) versus $\varepsilon$ at 23 mK. The measured current is decomposed in an elastic (dashed) and an inelastic (dotted-dashed) part.

FIG. 2. (**A**) Measured current versus $\varepsilon$ for $T = 23$ to 300 mK. The current is measured for $eV_{SD} = 140$ $\mu$eV while sweeping $V_{GR}$ and $V_{GL}$ simultaneously in opposite directions such that we change the energy difference $\varepsilon$. Gate voltage is translated to energy $\varepsilon$ by a calibration better than 10% using photon-assisted tunneling measurements (16). Dashed lines indicate exponential dependence, $e^{\varepsilon/kT}$, for $|\varepsilon| \gg kT$. Arrows point at step-like structure on the emission side ($\varepsilon > 0$) and a shoulder on the absorption side ($\varepsilon < 0$). From fits (11) to the elastic current part at 23 mK we obtain $h\Gamma_L \approx hT_c \approx 1$ $\mu$eV and $h\Gamma_R \approx 0.1$ $\mu$eV for this dataset. (**B**) Reconstructed current for different $T$. The spontaneous emission spectrum derived from the measured data at 23 mK and Eqs. 1 are used to reconstruct the full temperature and energy dependence. (**C**) The absorption rate $W_a$ (open symbols) and emission rate $W_e$ (closed symbols) normalized by the spontaneous emission rate $A$ versus $kT/|\varepsilon|$. Circles, squares, upper- and lower-triangles, and diamonds are taken at $|\varepsilon| = 18, 24, 40, 60,$ and 80 $\mu$eV, respectively (see also symbols in **A**). The solid line indicates the Bose-Einstein distribution, $\langle n \rangle$, whereas the dashed line shows $\langle n \rangle + 1$.

FIG. 3. Current spectrum for different coupling energies at 23 mK. (**A**) The magnetic field is 1.6 T for (i) and 2.4 T for the other curves (10). The curves have an offset, and curve (i) is multiplied by 5. Rough estimates for the coupling energies are: (i) $hT_c$ ($\sim 0.1$ $\mu$eV) $\ll h\Gamma_R$ ($\sim 10$ $\mu$eV), (ii) $hT_c$ ($\sim 1$ $\mu$eV) $\approx h\Gamma_R$ ($\sim 1$ $\mu$eV), (iii) $hT_c > h\Gamma_R$ ($\sim 0.1$ $\mu$eV), and (iv) $hT_c \gg h\Gamma_R$ ($\sim 0.01$ $\mu$eV) and $\Gamma_L \gtrsim \Gamma_R$ for all curves. The two dotted curves are the derivatives $-dI/d\varepsilon$ in arbitrary units for curves (i) and (ii) to enhance the bump-structure. (**B**) Logarithmic-linear plots for (i) and (ii). Dashed lines are Lorentzian fits. For (ii) we chose parameters that fit the tail for negative $\varepsilon$. (**C**) Logarithmic-logarithmic plots of the emission spectrum for two different samples. The sold lines are taken on the FIB sample in Fig. 1**A** (upper trace is the same as (ii) in A; lower trace is for coupling energies between (i) and (ii) in A). The dotted line is taken on a surface gate sample with a distance between left and right barriers of 600 nm (12). The dashed lines indicate a $1/\varepsilon$ and $1/\varepsilon^2$ dependence expected for piezo-electric interaction with 3D and 2D phonons, respectively.

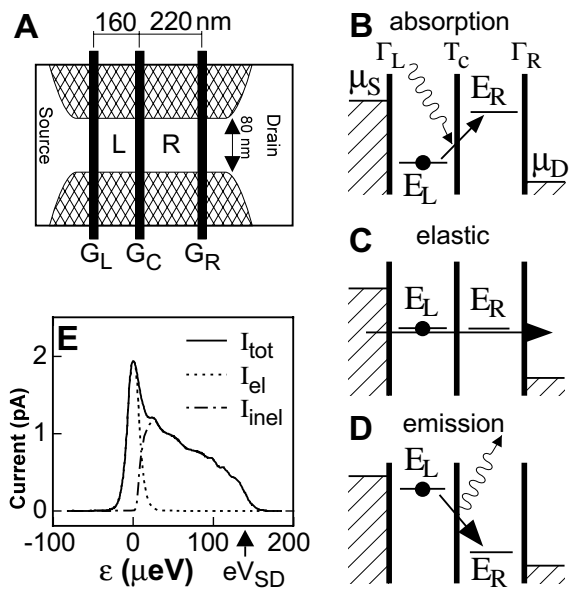

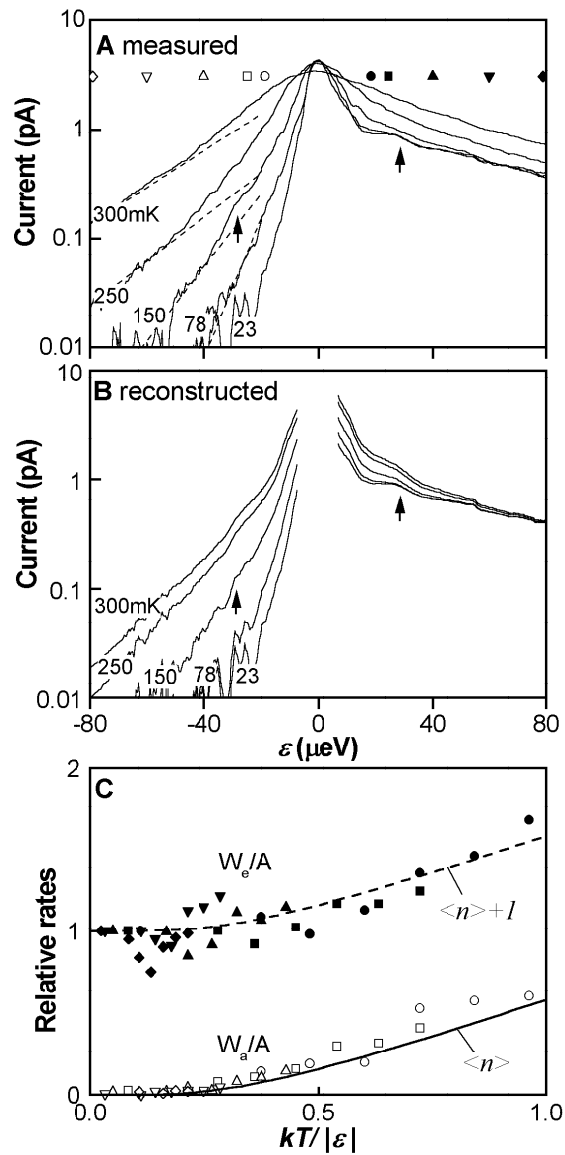

Fig. 1

Fig. 2

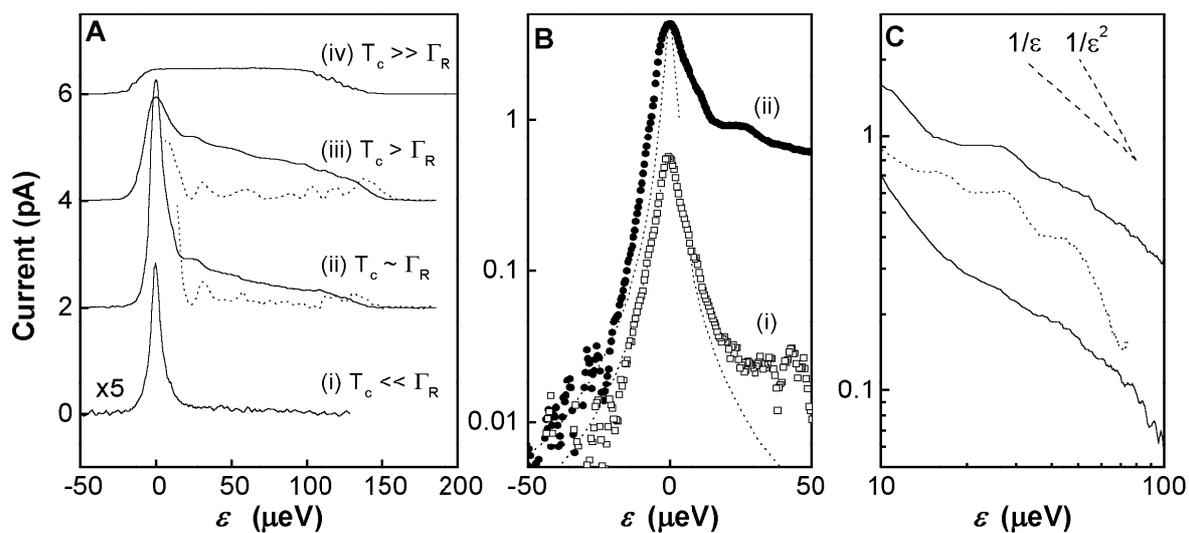

Fig. 3